\documentclass[journal,10pt]{IEEEtran}
\usepackage[utf8]{inputenc}
\usepackage{graphicx}
\usepackage{amsmath,amssymb}
\usepackage{float}
\usepackage{subcaption}
\usepackage{booktabs}
\usepackage{enumitem}
\usepackage[framemethod=TikZ]{mdframed}
\usepackage{soul}
\usepackage{advdate} 
\usepackage[ruled,vlined]{algorithm2e}
\usepackage[most]{tcolorbox}
\usepackage{color,xcolor}
\usepackage{setspace}
\usepackage{dblfloatfix}
\usepackage{capt-of} %
\usepackage{wrapfig}
\usepackage{lipsum}  % Optional: for dummy text like \lipsum[1]
\usepackage{array}

\usepackage{adjustbox}

\usepackage{amsmath}
\usepackage{nomencl}
\usepackage{cite}

\makenomenclature

\usepackage{hyperref}
\hypersetup{
    colorlinks=true,
    linkcolor=cyan,      % for \ref, \Cref, etc.
    citecolor=cyan,      % for \cite
    urlcolor=cyan        % for \url, \href
}

\usepackage[capitalise,noabbrev]{cleveref}  % Load AFTER hyperref

\crefname{equation}{equation}{equations}
\Crefname{equation}{Equation}{Equations}

\title{Physics-Informed Unit Commitment Framework for Nuclear Reactors}
\author{
    \IEEEauthorblockN{
        Shiny Choudhury\IEEEauthorrefmark{1},
        Michael Davidson\IEEEauthorrefmark{1}\IEEEauthorrefmark{2},
        George Tynan\IEEEauthorrefmark{1}
    }\\
    \thanks{Shiny Choudhury, Michael Davidson, and George Tynan
are with the Department of Mechanical and Aerospace Engineering, University of California at San Diego, La Jolla, CA 92093 USA (e-mail:
schoudhury@ucsd.edu; mrdavidson@ucsd.edu; gtynan@ucsd.edu).}
    \IEEEauthorblockA{\IEEEauthorrefmark{1}Mechanical and Aerospace Engineering, University of California, San Diego, CA, USA}\\
    \IEEEauthorblockA{\IEEEauthorrefmark{2}School of Global Policy and Strategy, University of California, San Diego, CA, USA}
}

\begin{document}
\newcommand{\red}{\textcolor{red}}
\newcommand{\SC}[1]{\red{SC: #1}}
\newcommand{\blue}{\textcolor{blue}}
\newcommand{\MD}[1]{\blue{MD: #1}}
\newcommand{\cyan}{\textcolor{cyan}}
\newcommand{\GT}[1]{\cyan{GT: #1}}
\definecolor{navy}{RGB}{0,0,128}
\definecolor{bluenavy}{HTML}{2B4A78}
\renewcommand\IEEEkeywordsname{Keywords}

% \doublespacing

\maketitle

\begin{abstract}
Nuclear reactors are often modeled as inflexible baseload generators with fixed downtimes and restrictive ramping constraints. In practice, however, a reactor’s operational flexibility is closely tied to its fuel cycle and associated reactivity margin. A key physical constraint for power maneuverability is \textit{xenon poisoning}, caused from the transient buildup of neutron‐absorbing xenon following a power reduction. This transient can delay or prevent subsequent power ramp-up due to suppressed core reactivity. Additionally, if a reactor is shutdown during periods of low reactivity, restart times can vary significantly, leading to prolonged downtimes. This work introduces a physics-informed modeling framework that embeds fuel cycle dynamics within a unit commitment (UC) formulation. The framework tracks reactivity margin, dynamically enforces xenon induced constraints, and endogenously schedules refueling outages based on core conditions. By capturing intracycle reactivity evolution, the model enables operation dependent nuclear dispatch that reflects both techno-economic requirements and irreducible nuclear physics limits. Application to a representative reactor fleet shows that flexible operation can slow reactivity degradation and extend fuel cycles. Results further demonstrate that different operational modes substantially affect VRE utilization, curtailment, and nuclear fleet capacity factors. These findings highlight the importance of fuel cycle aware flexibility modeling for accurate reactor scheduling and integration of nuclear power into energy system models.

\end{abstract}

\printnomenclature

\section{Introduction}
Nuclear power plants (NPPs) have traditionally been among the most inflexible thermal generators, but now face tremendous pressure to operate more flexibly due to the rapid integration of variable renewable energy (VRE) into electric grids~\cite{haratyk_early_2017}. With growing VRE penetration the variability of net load curve increases, placing rising demands on conventional generators to provide ramping capability, part load operation, and fast startup/shutdown services~\cite{impram_challenges_2020, schill_start-up_2017, gonzalez-salazar_review_2018, shaker_impacts_2016, lund_review_2015}. And since VRE increasingly sets marginal prices, NPPs must depart from their historical “baseload” operation mode and engage in levels of cycling and modulation rarely required in the past~\cite{jenkins_benefits_2018}. This shift challenges both the physical capabilities and economic assumptions that have traditionally underpinned nuclear operations. Despite these changing operational demands, large-scale power system models continue to represent NPPs as inflexible baseload units governed by static single-valued parameters~\cite{stauff_reactor_2021}. However, in reality, nuclear reactors exhibit fuel cycle dependent flexibility. For much of the fuel cycle, reactors can operate flexibly in response to grid signals, but as they approach the end of the cycle, core physics introduces additional binding constraints. These arise from the accumulation of fission fragments and neutron poisons (such as xenon-135) in the control rods, leading to increasing minimum generation levels and longer downtimes~\cite{franceschini_advanced_2008, lamarsh_introduction_2001}.

While a growing body of literature has started recognizing the importance of incorporating nuclear reactor physics and fuel cycle effects into power system models, most studies still assume limited or infrequent cycling of reactors, and treat operational constraints as fixed throughout the simulation horizon~\cite{ponciroli_profitability_2017, jenkins_benefits_2018, rahman_steady-state_2025, lynch_nuclear_2022}. As a result, most models decouple power output from the evolving physical state of the core. This abstraction carries significant consequences: because nuclear fuel is stored onboard, prolonged low capacity factor operation (as demanded in VRE-heavy mixes) extends the cycle length and postpones the onset of reactivity limiting conditions~\cite{alhadhrami_dispatch_2023}. And static NPP model's fail to capture such relationships.

This study addresses the modeling gap by introducing a physics informed unit commitment framework that explicitly links core reactivity margin, resulting xenon transient, and economic dispatch. Thus enabling a realistic representation of NPP behavior across different operational settings while capturing physical constraints. To the authors’ knowledge, this is the first tractable formulation that embeds full cycle nuclear physics in a unit commitment setting. By modeling these dynamic feedbacks, the framework moves beyond static constraint formulations and provides a stronger basis for assessing the role of NPPs in VRE dominated grids and otherwise.

\section{Framework}
Nuclear reactors operate under a complex set of constraints stemming from regulatory requirements, engineering design limits, and the fundamental physics of fission reaction. Operational constraints such as ramp rates, minimum generation levels, and downtime durations are typically imposed exogenously during standard dispatch~\cite{loflin_advanced_2014, stauff_reactor_2021}. For example, in the United States, a Westinghouse AP1000 reactor is typically permitted to reduce power from 100\% to 50\% and return to full power only once within a 24-hour window~\cite{noauthor_ap1000_nodate}. However, as the reactor progresses through its fuel cycle, physics driven constraints such as increasing minimum generation levels and longer downtime begin to apply. These constraints introduce a strong temporal dependence and reflect the evolving reactor core conditions. In this study, we first derive the physics-induced constraints as a function of fuel cycle progression and encode them into a precomputed lookup table. This table is used to dynamically enforce reactor level operational limits during the dispatch of a nuclear fleet in different scenarios. 

The widely deployed Westinghouse AP1000 Pressurized Water Reactor (PWR) serves as the reference technology for illustration of the formalism~\cite{noauthor_ap1000_nodate}. The AP1000 has a net electrical output of approximately 1000~MWe, and its primary loop uses pressurized water as both coolant and neutron moderator. Heat from this loop is transferred to a secondary circuit that operates
a Rankine cycle to generate electricity. The reactor core is composed of fuel assemblies arranged in a 17~$\times$~17 lattice, using low-enriched uranium dioxide (UO$_2$) with enrichment levels up to 5\% U-235. The remainder of this paper is organized as follows:

% Core reactivity is controlled via a combination of black and grey control rods made of boron carbide (B$_4$C), actuated through the control rod drive mechanism (CRDM)~\cite{noauthor_design_nodate}.
\begin{itemize}
    \item \Cref{xenon-poisoning} investigates core reactivity under various power ramping scenarios, focusing on the introduction of xenon poisoning and the resulting negative reactivity. 
    \item \Cref{core-reactivity-flex-power} presents a core reactivity margin degradation model adapted for flexible reactor power output.
    \item \Cref{xenon-poisoning-parameters} derives core reactivity dependent operational parameters, including limits on minimum generation level and downtime following a reactor shutdown.
    \item \Cref{unit-commitment} integrates these constraints in a modified unit commitment (UC) formalism, embedding fuel cycle tracking. This section introduces novel constraints to the canonical UC model for nuclear representation.
    \item \Cref{psudocode} presents the complete dispatch formalism in a structured algorithm.
\end{itemize}

\begin{figure*}[ht]
    \centering
    \subfloat[Ramp down \label{fig:ramp1}]{
        \includegraphics[scale=0.28]{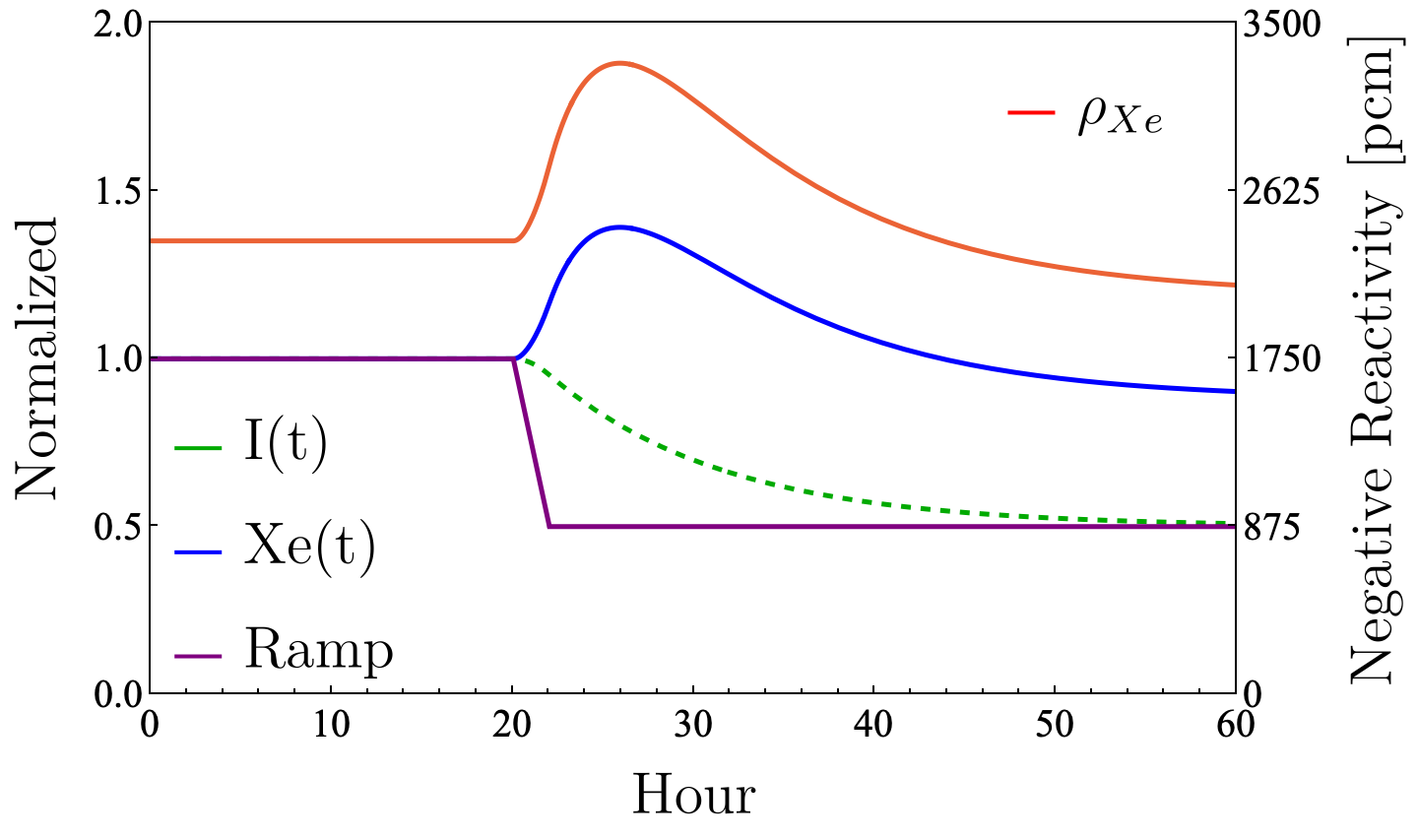}
    }
    \hspace{5mm}
    \subfloat[Ramp down and up\label{fig:ramp3}]{
        \includegraphics[scale=0.28]{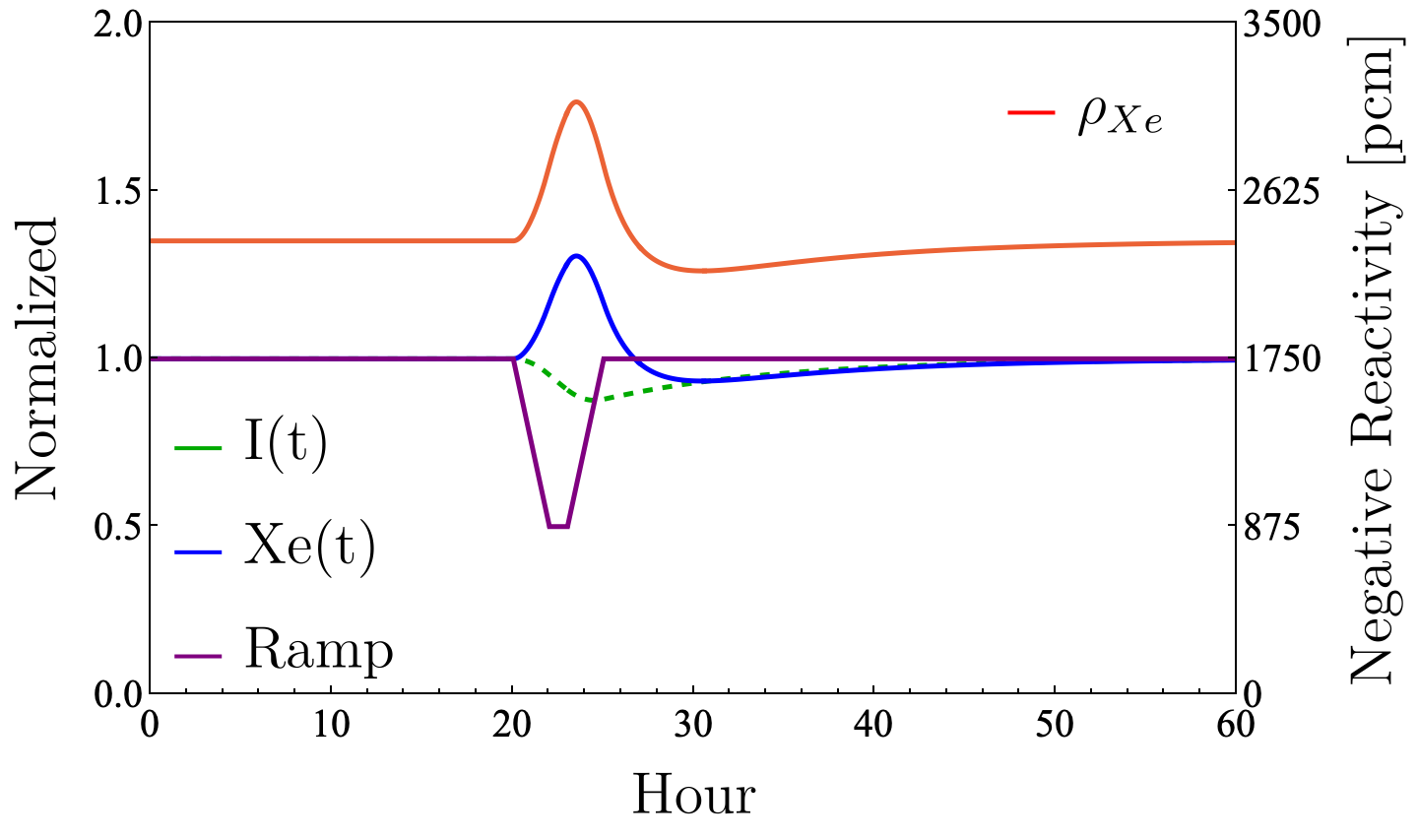}
    }
    \caption{\textsuperscript{135}Xe, \textsuperscript{135}I and negative reactivity ($\rho_{Xe}$) evolution for two different ramp scenarios.}
    \label{fig:reactivity}
\end{figure*}

\section{Determination of reactivity dependent parameters}
\subsection{Xenon poisoning and negative reactivity}\label{xenon-poisoning}
During a power ramp down in a nuclear reactor, the concentration of xenon (\textsuperscript{135}Xe) increases due to the radioactive decay of iodine (\textsuperscript{135}I), a fission product. \textsuperscript{135}Xe is a potent neutron absorber, and its concentration continues to rise for several hours after a downward ramp has ended~\cite{lamarsh_introduction_2001}. Within the reactor core, maintaining criticality requires precise neutron balance, which the \textsuperscript{135}Xe isotope disrupts. This effect, known as \textit{xenon poisoning}, reduces the core reactivity and temporarily limits the reactor's ability to ramp power back up. \Cref{eq:I} and \Cref{eq:Xe} describe the one-point differential equations governing the evolution of \textsuperscript{135}I and \textsuperscript{135}Xe concentrations as a function of the core neutron flux~\cite{lamarsh_introduction_2001}. The time varying neutron flux, $\phi(t)$, is modeled as proportional to the reactor power $P(t)$, as shown in \Cref{eq:Phi}.
 
\begin{equation}
\frac{d\,{}^{135}\mathrm{I}(t)}{dt} = -\lambda_I \, {}^{135}\mathrm{I}(t) + \gamma_I \, \phi(t) \, \bar{\Sigma}_f
\label{eq:I}
\nomenclature{${}^{135}\mathrm{I}(t)$}{Instantaneous concentration (atoms/cm\textsuperscript{3}) of \textsuperscript{135}I at time $t$}
\nomenclature{${}^{135}\mathrm{Xe}(t)$}{Instantaneous concentration (atoms/cm\textsuperscript{3}) of \textsuperscript{135}Xe at time $t$}
\nomenclature{$\lambda_I$}{Decay constant of \textsuperscript{135}I}
\nomenclature{$\gamma_I$}{Effective yield of \textsuperscript{135}I}
\nomenclature{$\bar{\Sigma}_f$}{Macroscopic fission cross-section}
\nomenclature{$\phi (t)$}{Instantaneous nuclear flux (neutrons/cm$^2$-hr)}
\nomenclature{$\phi_0$}{Average nuclear flux (neutrons/cm$^2$-hr)}
\nomenclature{$P(t)$}{Reactor power output at time $t$ in MW}
\end{equation}

\begin{align}
\frac{d\,{}^{135}\mathrm{Xe}(t)}{dt} &= \lambda_I \, {}^{135}\mathrm{I}(t) - \lambda_{Xe} \, {}^{135}\mathrm{Xe}(t) \nonumber\\
&\quad +\, \gamma_{Xe} \, \phi(t) \, \bar{\Sigma}_f 
- \sigma_{abs}^{Xe} \, \phi(t) \, {}^{135}\mathrm{Xe}(t)
\label{eq:Xe}
\nomenclature{$\lambda_{Xe}$}{Decay constant of \textsuperscript{135}Xe}
\nomenclature{$\gamma_{Xe}$}{Effective yield of \textsuperscript{135}Xe}
\nomenclature{$\sigma_{abs}^{Xe}$}{Microscopic absorption cross-section of \textsuperscript{135}Xe}
\end{align}

\begin{equation}
\phi(t) = \phi_0 \times \left( \frac{P(t)}{P_{\max}} \right)
\label{eq:Phi}
\end{equation}

$\phi_0$ is the average neutron flux in the core, expressed in units of $\text{neutrons}/\text{cm}^2\text{-hr}$. $\lambda_I$ and $\lambda_{Xe}$ are the decay constants of \textsuperscript{135}I and \textsuperscript{135}Xe, respectively, with units of $\text{hr}^{-1}$. $\gamma_I$ and $\gamma_{Xe}$ represent the effective yields of \textsuperscript{135}I and \textsuperscript{135}Xe per fission. $\sigma_{abs}^{Xe}$ denotes the microscopic absorption cross section of \textsuperscript{135}Xe. The values for all parameters are summarized in \Cref{tab:parameter}~\cite{ponciroli_profitability_2017, franceschini_advanced_2008}. 

The steady state concentrations of \textsuperscript{135}I and \textsuperscript{135}Xe are obtained by setting the time derivatives in \Cref{eq:I} and \Cref{eq:Xe} to zero and solving for \textsuperscript{135}$\mathrm{I}_{\text{eq}}$ and \textsuperscript{135}$\mathrm{Xe}_{\text{eq}}$, respectively. The derived expressions are shown in \Cref{eq:Ieq} and \Cref{eq:Xe_eq}.

\begin{equation}
{}^{135}\mathrm{I}_{\text{eq}} = \frac{\gamma_I \, \phi_0 \, \bar{\Sigma}_f}{\lambda_I}
\label{eq:Ieq}
\nomenclature{${}^{135}\mathrm{I}_{\text{eq}}$}{Equilibrium concentration of \textsuperscript{135}I}
\end{equation}

\begin{equation}
{}^{135}\mathrm{Xe}_{\text{eq}} = \frac{ \phi_0 \, \bar{\Sigma}_f \left( \gamma_I + \gamma_{Xe} \right) }{ \lambda_{Xe} + \sigma_{abs}^{Xe} \, \phi_0 }
\label{eq:Xe_eq}
\nomenclature{${}^{135}\mathrm{Xe}_{\text{eq}}$}{Equilibrium concentration of \textsuperscript{135}Xe}
\end{equation}

\textsuperscript{135}Xe absorbs the fission neutrons and introduces a negative reactivity within the core (also referred to as `Xenon defect') that can be computed using \Cref{eq:rho_Xe}. 

\begin{equation}
\rho_{Xe}(t) \approx \frac{ \sigma_{abs}^{Xe} \, {}^{135}\mathrm{Xe}(t) }{ v \, \bar{\Sigma}_f }
\label{eq:rho_Xe}
\nomenclature{$\rho_{Xe}(t)$}{Negative reactivity or Xenon defect at time $t$ in pcm}
\nomenclature{$v$}{Average number of neutrons produced per fission}
\end{equation}

In \Cref{fig:reactivity}. the evolution of \textsuperscript{135}I, \textsuperscript{135}Xe, and the resulting negative reactivity for two distinct power ramp scenarios are shown. During a power ramp down, the \textsuperscript{135}Xe concentration transiently increases due to \textsuperscript{135}I decay before eventually decreasing and equilibrating (\cref{fig:ramp1}). While during a power ramp up immediately following a ramp down event (\cref{fig:ramp3}), we observe that the peak negative reactivity is smaller in magnitude compared to an isolated ramp down, and the transients stabilizes more quickly. This peak depends on (1) the power level before the ramp down begins and (2) the depth of the power reduction or the $\Delta P$~\cite{franceschini_advanced_2008, ponciroli_profitability_2017, lamarsh_introduction_2001}

% In this study we want to avoid negative reactivity induced power stagnation altogether, hence we are interested in the peak values of the xenon defect in various ramping events. .

% Conversely, during a power ramp up, \textsuperscript{135}Xe concentration initially decreases before stabilizing at a new higher equilibrium value corresponding to the higher power level (\cref{fig:ramp2}). The resulting xenon defect in both these ramping scenarios follow the trend of \textsuperscript{135}Xe evolution.

\subsection{Core reactivity margin with flexible power output}\label{core-reactivity-flex-power}
For PWRs, flexibility can be realized by ramping core power or by venting steam before it reaches the turbine~\cite{ingersoll_can_2015}. Core ramping is typically performed using control rods or by varying the boron concentration in the coolant (referred to as \emph{chemical shim})~\cite{franceschini_advanced_2008}. The control rod assembly includes a combination of highly absorbent `black rods' and partially absorbent `grey rods', each characterized by different neutron absorption cross sections. By controlling the insertion and withdrawal of these rods, operators can regulate neutron flux and thereby control reactor power. However, over time, neutron absorbing fission products accumulate within fuel rods, reducing the core reactivity margin and narrowing the range of achievable power maneuvers~\cite{ponciroli_profitability_2017}. During these periods, a severe case of xenon poisoning can lead to power stagnation for several hours.

Consequently, the operational flexibility achievable at any given time is a balance between the core reactivity margin (determined by the operational history of the fuel) and the gradually increasing impact of negative reactivity during a large power ramp down. To quantify core reactivity, we use the effective multiplication factor, $\mathrm{k}_{\mathrm{eff}}$, that measures the ratio of the neutron population from one generation to the next. When freshly fueled, a reactor has $\mathrm{k}_{\mathrm{eff}} > 1$ to sustain criticality despite fuel burnup and other losses. At steady state, an instantaneous $\mathrm{k}_{\mathrm{eff}} =1$ is maintained at all times by the use of control rods and/or chemical shims~\cite{franceschini_advanced_2008}. The remaining core reactivity margin\footnote{Reactivity is expressed in pcm (per cent mille), where 1 pcm = $1\times10^{-5}$} at different stages of the fuel cycle (indexed by $n$) is computed as shown in \Cref{eq:margin}. 

\begin{equation}
    \Delta \rho_{\operatorname{marg, n}}=\frac{\mathrm{k}_{\mathrm{eff, n}}-1}{\mathrm{k}_{\mathrm{eff, n}}} \times 10^5\ \text{[pcm]}
    \nomenclature{$\Delta \rho_{\operatorname{marg, n}}$}{Reactivity Margin for iteration $n$}
    \label{eq:margin}
    \nomenclature{$k_{eff, n}$}{Effective multiplication factor for iteration $n$}
    \nomenclature{$\mathrm{\alpha}_{\mathrm{n,g}}$}{Capacity factor for iteration $n$ and generator $g$}
    \nomenclature{$P_0$}{Stable power level after ramp down ends in MW}
\end{equation}

% Note that the value $k_{eff,n}$ is the multiplication factor in the absence of control rod and chemical shim and thus takes value $\geq 1$. The parameter $k_{eff,n}$ starts at an initial value at beginning-of-life (BOL), denoted $k_{eff,BOL}$, and gradually degrades as fissile nuclei are consumed, eventually approaching a minimum of $k_{eff,n} \simeq 1$, at which point the reactor must undergo refueling.

For nuclear reactors, `fuel burnup' quantifies the cumulative energy extracted from the nuclear fuel over time. It is a useful metric because a freshly fueled PWR core contains enough fissile material to support a range of burnup trajectories, depending on the reactor's time-varying capacity factor during operation. At full power operation, the reactor undergoes maximum burnup, and the degradation rate \( m \) is approximated from empirical observations as \( 3.8 \times 10^{-4}~\text{day}^{-1} \)~\cite{franceschini_advanced_2008}. During flexible power output, $\mathrm{k}_{\mathrm{eff}}$ degradation will be dictated by the magnitude of part load operation, which will then be reflected in the burnup levels. To model this relationship, we define a simple recursive relation in \Cref{eq:recursive_reln}, where the degradation rate $m$ is scaled by the capacity factor $\alpha_n$ to compute the effective multiplication factor $k_{\mathrm{eff,n+1}}$ using $k_{\mathrm{eff,n}}$. The capacity factor $\alpha_\mathrm{n}$ (defined in \Cref{eq:ALPHAn}), is the ratio of actual generation to the theoretical maximum over period $T$. The updated value $k_{\mathrm{eff,n+1}}$ is then used to calculate the remaining reactivity margin, which directly affects the reactor's operational flexibility in period $\mathrm{n+1}$.

\begin{equation}
    \mathrm{k}_{\mathrm{eff}, \mathrm{n}+1}=\mathrm{k}_{\mathrm{eff}, \mathrm{n}}-m \times \alpha_{\mathrm{n}}
    \label{eq:recursive_reln}
\end{equation}

\begin{equation}
    \alpha_n = \frac{\sum_{t=1}^{T} P(t)}{P_{\max} \times T}
    \label{eq:ALPHAn}
\end{equation}

\begin{figure}
    \centering
    \includegraphics[scale=0.3]{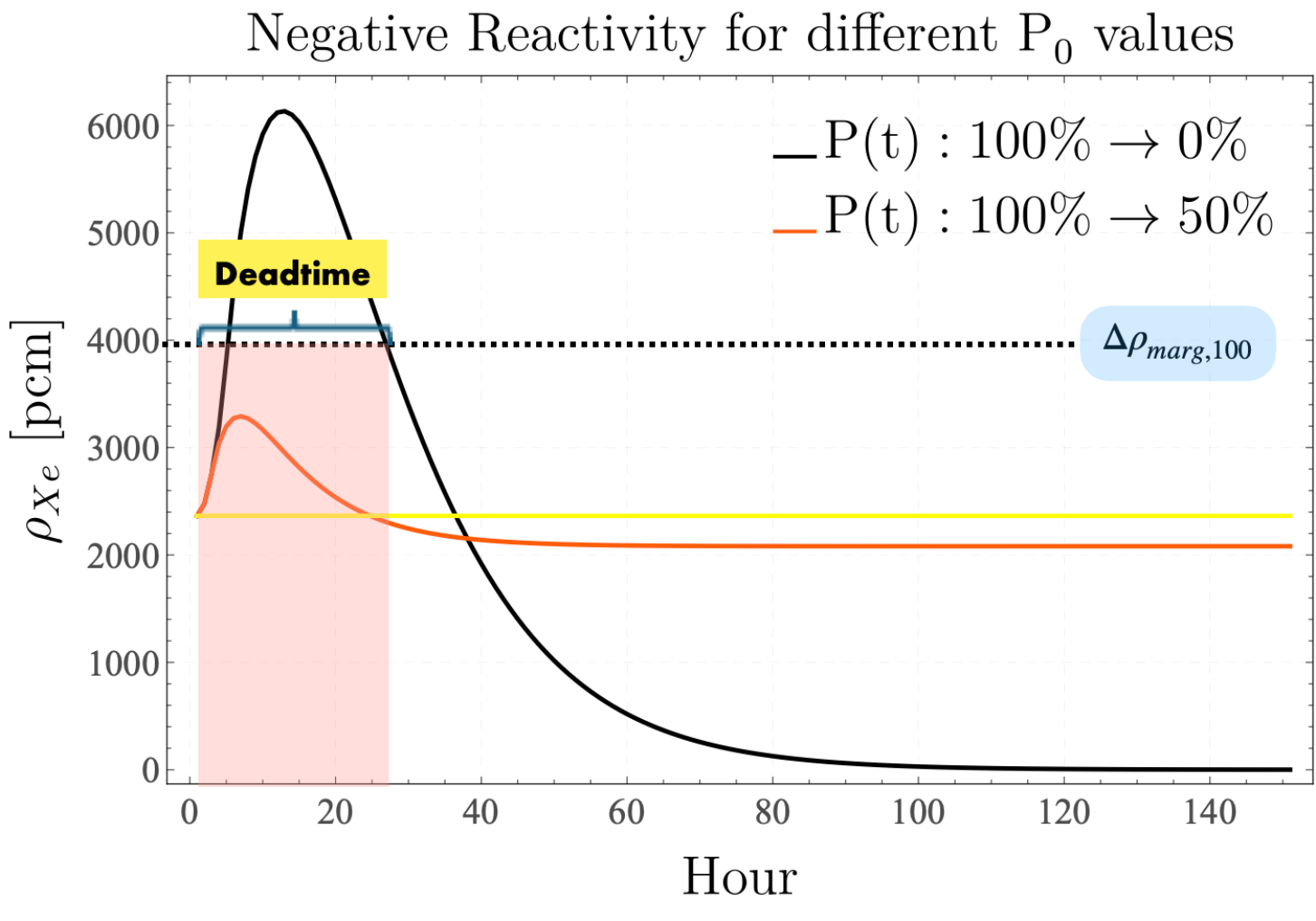}
    \caption{Deadtime and minimum generation level are determined by the evolution of $\rho_{Xe}$ following a power reduction from full power, relative to the available reactivity margin. For illustration, we consider ramps of $P_{\max} \rightarrow 0.5P_{\max}$ and $P_{\max} \rightarrow 0$. Deadtime persists as long as $\rho_{Xe} > \Delta \rho_{\text{margin},100}$. The minimum generation level corresponds to the $P_0$ at which the peak negative reactivity remains within the available margin; in this example, $P_0 = 0.5P_{\max}$.}
    \label{fig:deadtime_comp}
\end{figure}

\subsection{Reactivity margin dependent parameters}\label{xenon-poisoning-parameters}
As a reactor approaches low reactivity levels due to fuel burnup, limits on allowable ramps begin to apply. The minimum generation levels gradually increases until the reactor can offer no ramping without facing a power stagnation~\cite{franceschini_advanced_2008, ponciroli_profitability_2017, jenkins_benefits_2018}. Additionally, longer downtime start applying in case of a shutdown \cite{lamarsh_introduction_2001}. In the following section we compute these parameters which are then used as inputs in a UC model.

\subsubsection{Minimum generation level}\label{min_power_level}
To maximize operational flexibility, xenon poisoning-induced power stagnation must be avoided at all stages of core reactivity. As a consequence, the accessible minimum generation level increases slowly as the reactor progresses in its fuel cycle. To compute these minimum generation levels, we start by defining a piecewise ramp down from full power ($\mathrm{P_{\max}}$) to a parameterized value $\mathrm{P_{0}}$ at a ramp rate of 25\%/hr, as shown in \Cref{eq:piecewise_ramp}~\cite{stauff_reactor_2021}. $n_{0}$ is the hour at which the ramp down ends when power stabilizes at $\mathrm{P_0}$.

\begin{equation}
\mathrm{P}(t)=
\begin{cases}
\mathrm{P}_{\max}, & t = 0 \\
\mathrm{P}_{\max} \times \left(1 - 0.25\, t\right), & 0 < t \leq n_0 \\
\mathrm{P}_0, & t > n_0
\end{cases}
\label{eq:piecewise_ramp}
\end{equation}

This piecewise ramp down function is used to approximate the neutron flux in the core using \Cref{eq:Phi}, which is then used to compute the xenon defect evolution via \Cref{eq:rho_Xe}. The peak negative reactivity, denoted by \( \rho_{\mathrm{Xe,\max}(P_0)} \), is evaluated over a range of \( \mathrm{P_0} \) values, where power ramps from \( \mathrm{P_{\max} \rightarrow P_0} \) are precomputed and stored in a lookup table. For a given stage in the fuel cycle $n$, the minimum allowable generation level or $P_{\min,n}^*$ is determined based on the remaining core reactivity margin, $\Delta \rho_{\text{marg},n}$, as defined in \Cref{eq:Pmin_selection}. $P_{\min,n}^*$ represents the lowest accessible generation level while avoiding xenon poisoning induced power stagnation.

\begin{equation}
\begin{aligned}
P_{\min,n}^{\ast} = \operatorname*{arg\,min}_{P_{0}} \quad & \Delta \rho_{marg,n} - \rho_{Xe,\max}(P_{0}) \\
\text{s.t.} \quad & \Delta \rho_{marg,n} - \rho_{Xe,\max}(P_{0}) > 0
\end{aligned}
\label{eq:Pmin_selection}
\nomenclature{$\rho_{Xe,\max}(P_{0})$}{Peak negative reactivity for ramp down from 100\% power to $P_0$ in pcm}
\end{equation}

\begin{figure}
    \centering
    \includegraphics[scale=0.55]{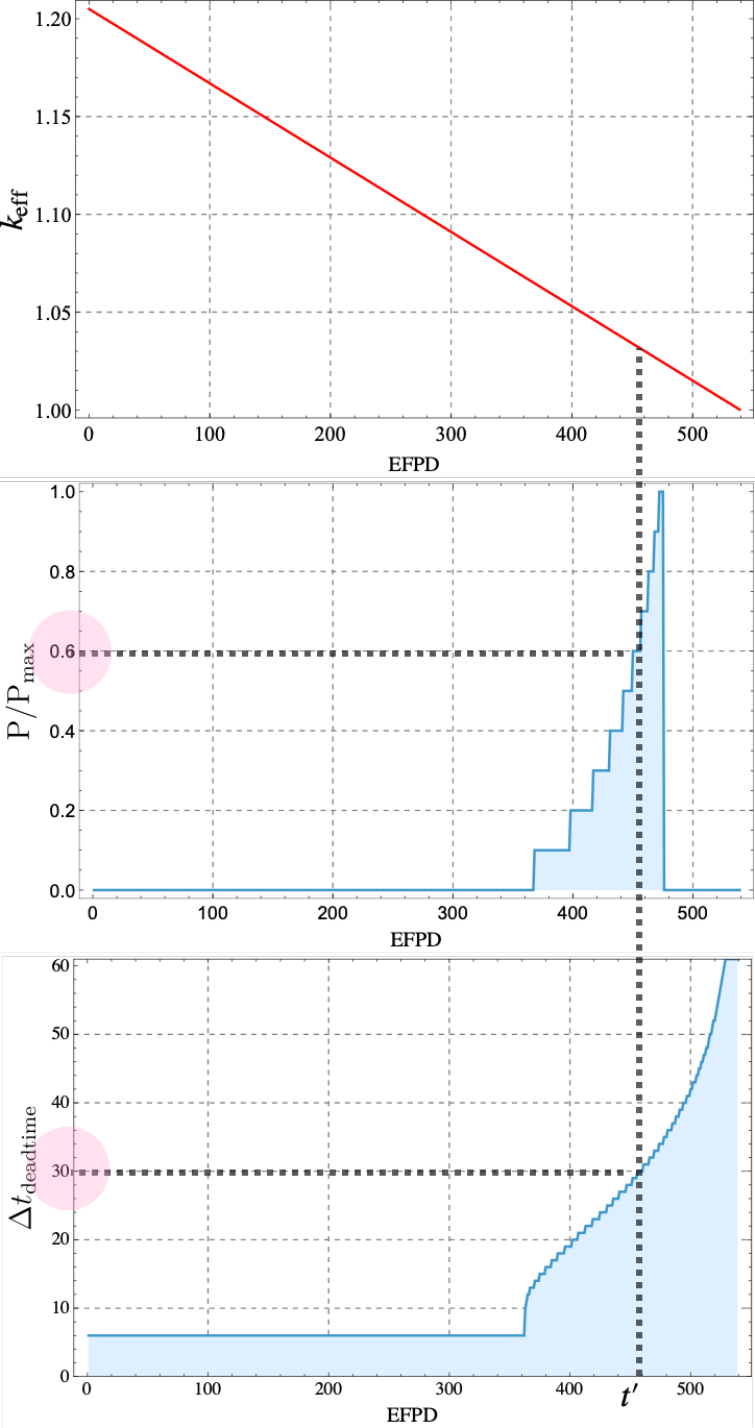}
    \caption{Determination of exogenously set $P_{min}$ and Deadtime value based on the core reactivity before short term UC dispatch.}
    \label{fig:pmin_deadtime_keff}
\end{figure}

\subsubsection{Deadtime after shutdown}\label{deadtime_after_shutdown}
In rare instances, a reactor may need to shutdown when the core reactivity is low. Such a shutdown can be treated as a power ramp down to zero and introduces additional constraints for subsequent restarts. In such a case, without an ongoing neutron flux, the \textsuperscript{135}Xe concentration transiently increases to substantially higher peaks since there are no active \textsuperscript{135}Xe sinks~\cite{lamarsh_introduction_2001}. To safely restart a reactor, a mandatory downtime—referred to as \emph{deadtime}—must be observed to allow for significant \textsuperscript{135}Xe to decay. The duration of this deadtime, denoted by \( \Delta t_{\text{deadtime},n} \), is defined as the time interval following a shutdown after which the negative reactivity or \( \rho_{Xe}(t) \) is strictly less than the available reactivity margin, \( \Delta \rho_{\text{marg},n} \) and decreasing. After this period, the reactor may safely restart without violating core reactivity constraints. Assuming a full power to zero shutdown, the deadtime computation is given by \Cref{eq:deadtime_comp}.

\begin{equation}
\Delta t_{\text{deadtime},n} = \operatorname*{arg\,max}_{t}\left\{ t \in \mathbb{R}^+ \;\middle|\; \rho_{Xe}(t) > \Delta \rho_{\text{marg},n} \right\}
\label{eq:deadtime_comp}
\end{equation}

\Cref{fig:deadtime_comp}. illustrates the evolution of the negative reactivity following a shutdown from full-power operation, the available reactivity margin on day 100, and the resulting deadtime that the reactor must observe before restart is feasible. In the shown figure, the deadtime is at least 26 hours, after which the core has enough excess reactivity to override the negative reactivity and permit a restart. A precomputed deadtime value table is derived using \Cref{eq:deadtime_comp} for various reactivity margins.

% \subsubsection{Reactivity dependent constraint} 
Exhaustive xenon transient simulations for different ramp sequences are performed to generate a lookup table that links minimum allowable power level ($P_{\min,n}^*$) and post shutdown deadtime ($\Delta t_{deadtime,n}$) to the remaining reactivity margin. \Cref{fig:pmin_deadtime_keff}. illustrates pictorially the lookup table parameter enforcement across different $\mathrm{k_{eff}}$ values as the fuel undergoes burnup. In effect, we adjust reactivity degradation based on reactor operation at full load or part load by scaling degradation based on burnup. This allows us to track reactivity decline over time, impose appropriate operating constraints, and prevent power stagnation. 

\section{Unit Commitment with reactivity tracking}\label{unit-commitment}
Power plants operate under a range of technical constraints such as ramp up/down rates, minimum up and down times, and startup/shutdown durations—all of which must be considered when scheduling a heterogeneous fleet of generators. Unit Commitment (UC) is a mixed integer linear programming (MILP) formulation widely used for this purpose~\cite{knueven_mixed-integer_2020}. The UC model determines the optimal commitment or on/off status and generation levels for individual generators to minimize total operational cost while satisfying all technical constraints. We modify the UC model to incorporate inputs from a reactivity tracking algorithm, which imposes precomputed constraints from the lookup tables, enabling a realistic representation of NPP operation.

\subsection{Objective function}
The objective of the UC formulation includes the canonical components: variable generation cost, startup cost, and a penalty for non-served energy (NSE). Additionally, a shutdown cost term is included to capture the operational costs associated with shutting down nuclear reactors, as motivated in~\cite{stauff_reactor_2021}. The objective expression is described in \Cref{eq:obj_uc}.

\begin{align}
\min & \sum_{g \in G} C_{\text{var},g} \cdot p_{g,t} 
+ \sum_{t \in T} C_{nse} \cdot nse_t \nonumber \\
& + \sum_{g \in G_{\text{nuc}}} \sum_{t \in T}
\left( C_{\text{start},g} \cdot z^{\text{start}}_{g,t}
+ C_{\text{shut},g} \cdot z^{\text{shut}}_{g,t} \right)
\nomenclature{$T$}{Time horizon of dispatch}
\nomenclature{$G$}{Set of all generators}
\nomenclature{$G_{\text{nuc}}$}{Subset of generators that are nuclear reactors ($G_{\text{nuc}} \subset G$)}
\nomenclature{$C_{var,g}$}{Variable cost of generator $g$ (\$/MWh)}
\nomenclature{$C_{start,g}$}{Startup cost of generator $g$ (\$/MW)}
\nomenclature{$C_{shut,g}$}{Shutdown cost of generator $g$ (\$/MW)}
\nomenclature{$p_{g,t}$}{Generation from generator $g$ at time $t$ (MWh)}
\nomenclature{$C_{nse}$}{Non-served energy (NSE) penalty (\$/MWh)}
\nomenclature{$z^{start}_{g,t}$}{Binary startup indicator for generator $g$ at time $t$}
\nomenclature{$z^{shut}_{g,t}$}{Binary shutdown indicator for generator $g$ at time $t$}
\label{eq:obj_uc}
\end{align}

\subsection{Unit commitment constraints}
For discrete commitment of thermal generators, we use the three variable startup, shutdown and commit constraints~\cite{carrion_computationally_2006}. The generators are constrained by to minimum power levels, ramp rates, uptime and downtime. Additionally, we introduce two key changes for nuclear reactor representation--a variable minimum power level and a variable downtime based on remaining reactivity margins. \Cref{eq:min_gen_commit} to \Cref{eq:commit_transition} summarize the commitment constraints that discretely enforce minimum generation level and downtime on all thermal generators. $z^{\text{on}}_{g,t},\ z^{\text{start}}_{g,t},\ z^{\text{shut}}_{g,t} \in \{0,1\}$ are binary variables representing the commitment status, startup, and shutdown actions, respectively. 

\begin{align}
 p_{g,t} \geq P^{\min}_{g,n} \times z^{on}_{g,t}  &\quad \forall g \in G_{\text{nuc}},\ t \in T \label{eq:min_gen_commit} \\
 p_{g,t} \leq P^{\max}_g \times z^{on}_{g,t} &\quad \forall g \in G_{\text{nuc}},\ t \in T \label{eq:max_gen_commit} \\
 z^{on}_{g,t} \geq \sum_{t' = t - MinUp_g}^{t} z^{start}_{g,t'} &\quad \forall g \in G_{\text{nuc}},\ t \in T \label{eq:minup_logic} \\
 1 - z^{on}_{g,t} \geq \sum_{t' = t - MinDn_{g,n}}^{t} z^{shut}_{g,t'} &\quad \forall g \in G_{\text{nuc}},\ t \in T \label{eq:mindn_logic} \\
 z^{on}_{g,t+1} - z^{on}_{g,t} = z^{start}_{g,t+1} - z^{shut}_{g,t+1} &\quad \forall g \in G_{\text{nuc}},\ t \in T{-}1
\label{eq:commit_transition}
\end{align}

\nomenclature{$z^{on}_{g,t}$}{Binary commitment indicator for generator $g$ at time $t$}
\nomenclature{$P^{\min}_{g,n}$}{Minimum generation level for generator $g$ on dispatch block $n$}
\nomenclature{$P^{\max}_{g}$}{Maximum generation level for generator $g$}
\nomenclature{$MinDn_{g,n}$}{Minimum downtime following shutdown for generator $g$ on dispatch block $n$}
\nomenclature{$MinUp_{g}$}{Minimum uptime following commitment for generator $g$}

Typically, nuclear reactors are exogenously constrained to operate within a limited power range due to economic or regulatory requirements. Hence, the generation from each reactor must stay within 
\(
p_{g,t} \in \left[ P^{\min}_{g,n},\ P_{\max} \right]
\)
where the minimum generation level, $P^{\min}_{g,n}$ is defined as:
\[
P^{\min}_{g,n} = \max\left( P_{\min,\text{base}},\ P_{\min,n}^* \right)
\]

Here, $P_{\min,\text{base}}$ is an exogenously set minimum generation level, while $P_{\min,n}^*$ is the level determined by the remaining reactivity margin on day $n$ as computed in~\Cref{min_power_level}. Similarly, the minimum downtime for each reactor is defined as:
\[
\text{MinDn}_{g,n} = \max\left( \text{MinDn}_{\text{base}},\Delta t_{deadtime,n} \right)
\]
where $\text{MinDn}_{\text{base}}$ is the baseline downtime dictated by mechanical wear, Rankine cycle inertia, and safety protocols, and $\Delta t_{deadtime,n}$ accounts for xenon-induced deadtime, also precomputed based on reactivity levels on day $n$ as computed in~\Cref{deadtime_after_shutdown}.

\subsection{Minimum stable time constraints}
Following a power ramp down, nuclear reactors need to maintain a stable power level for a minimum duration before they can ramp power again. This is for operators to manage core conditions while accounting for xenon transients~\cite{noauthor_ap1000_nodate}. To capture this, we begin by introducing three binary variables \( rd_{g,t},\ st_{g,t},\ up_{g,t} \in \{0,1\} \), defined for each reactor \( g \in G_{\text{nuc}} \) and time step \( t \in T \), as originally formulated by Jenkins et al.~\cite{jenkins_benefits_2018}. The variable \( rd_{g,t} \) serves as a ramp down indicator and is activated (i.e., transitions from 0 to 1) when a ramp down occurs between \( t{-}1 \) and \( t \). Similarly, \( up_{g,t} \) indicates a ramp up over the same interval. Both ramp down and ramp up events are defined by a threshold change in generation by \( \delta \), as captured in \Cref{eq:delta_rampup_aux} and \Cref{eq:delta_rampdn_aux}. 

When a ramp down ends or \( rd_{g,t}\) transitions from 1 to 0 between $t-1$ to $t$, the stable state indicator \( st_{g,t}=1 \) is activated, requiring the reactor to remain at a constant generation level for \( \Delta t_{\text{stable}}\) periods before the next power ramp up is allowed. Finally, the sum of the three binary variables \( rd_{g,t}, up_{g,t}, st_{g,t} \) must equal 1 at each time step, and these binaries are only active when a unit is committed, i.e., \( z^{on}_{g,t} = 1 \). \Cref{eq:genaux_def} – \Cref{eq:commit_consistency} formalizes this logic and summarize the constraints.
In \Cref{tab:binary_schedule} state transition of the binary variables for an illustrative ramping scenario with a stability duration of \( \Delta t_{\text{stable}} = 2 \) hrs is shown.

\begin{align}
& p^{aux}_{g,t} = p_{g,t} - P^{\min}_{g,n} \times z^{on}_{g,t} \label{eq:genaux_def} \\
& p^{aux}_{g,t+1} - p^{aux}_{g,t} \geq \delta - M \times (1-up_{g,t})  \label{eq:delta_rampup_aux} \\
& p^{aux}_{g,t+1} - p^{aux}_{g,t} \leq up_{g,t} \times RampUp_g \label{eq:rampup_aux} \\
& p^{aux}_{g,t} - p^{aux}_{g,t+1} \geq \delta - M \times (1-rd_{g,t}) \label{eq:delta_rampdn_aux} \\
& p^{aux}_{g,t} - p^{aux}_{g,t+1} \leq rd_{g,t} \times RampDn_g \label{eq:rampdn_aux} \\
& \begin{aligned}
    st_{g,t+k} &\geq rd_{g,t} - rd_{g,t+1} \\
    &\text{\quad for all } t \in [1, T{-}\Delta t_{\text{stable}}{-}1], \\
    &\quad\quad\;\; k \in [1, \Delta t_{\text{stable}}], \\
    &\quad\quad\;\; g \in G_{\text{nuc}}
\end{aligned} \label{eq:pminstable_enforce} \\
& rd_{g,t} + up_{g,t} + st_{g,t} = z^{on}_{g,t} \label{eq:commit_consistency} \quad \forall g \in G_{\text{nuc}},\; t \in T
\end{align}
\nomenclature{$rd_{g,t}$}{Binary ramp down indicator for generator $g$ at time $t$}
\nomenclature{$up_{g,t}$}{Binary ramp up indicator for generator $g$ at time $t$}
\nomenclature{$st_{g,t}$}{Binary stable period indicator for generator $g$ at time $t$}
\nomenclature{$\Delta t_{deadtime,n}$}{Deadtime period for iteration $n$}
\nomenclature{$\Delta t_{stable}$}{Period of time when power transition is not allowed}
\nomenclature{$p^{aux}_{g,t}$}{Auxiliary variable for power output for generator $g$ at time $t$}

\begin{table}[ht]
\centering
\caption{Illustrative state transitions for minimum stable time implemtation with $\Delta t_{\text{stable}} = 2$.}
\label{tab:binary_schedule}
\begin{tabular}{c|cccc|p{3.5cm}}
\toprule
\textbf{$t$} & $\mathbf{z^{on}_{t}}$ & $\mathbf{up_{t}}$ & $\mathbf{rd_t}$ & $\mathbf{st_t}$ & \textbf{Description} \\
\midrule
0 & 1 & 0 & 0 & 1 & Initial steady-state generation \\
1 & 1 & 0 & 1 & 0 & Ramp-down begins \\
2 & 1 & 0 & 1 & 0 & Ramp-down continues \\
3 & 1 & 0 & 0 & 1 & Stability period enforced \\
4 & 1 & 0 & 0 & 1 & Stability period enforced \\
5 & 1 & 1 & 0 & 0 & Ramp-up begins \\
6 & 1 & 1 & 0 & 0 & Ramp-up continues \\
7 & 1 & 0 & 0 & 1 & Return to stable generation \\
8 & 0 & 0 & 0 & 0 & Reactor shut down \\
\bottomrule
\end{tabular}
\end{table}

\begin{algorithm}[!htb]
\caption{Modified Unit Commitment with Reactivity Tracking}
\label{algo:smr}
\KwIn{$k_{BOL,g}$, $P^{\min}_{g}$, downtime lookup tables, $\Delta t_{\text{refuel}}$}
\KwOut{$p_{g,t}$ $\forall g \in G_{\text{nuc}},\ t \in T$}

\ForEach{$g \in G_{\text{nuc}}$}{
    $k_{0,g} \gets k_{BOL,g}$ \\
    $\tau_g \gets 0$
}

\For{$n = 1$ \KwTo $N$}{
    \ForEach{$g \in G_{\text{nuc}}$}{
        \If{$k_{n,g} \leq 1$ \textbf{and} $\tau_g = 0$}{
            $\tau_g \gets \Delta t_{\text{refuel}}$
        }

        \If{$\tau_g > 0$}{
            Set $p_{g,t} \gets 0$ for all $t \in T$ \\
            $\tau_g \gets \tau_g - 1$ \\
            \If{$\tau_g = 0$}{
                $k_{n,g} \gets k_{BOL,g}$
            }
            \textbf{continue} \\
        }

        $\Delta \rho_{g,n}^{\text{marg}} \gets \frac{1 - k_{n,g}}{k_{n,g}} \cdot 10^5$\;
        Retrieve $P_{g,n}^{\min}$ and $MinDn_{g,n}$ from lookup table using $\Delta \rho_{g,n}^{\text{marg}}$ \\
    }

    Solve short-term UC \\

    \ForEach{$g \in G_{\text{nuc}}$}{
        $\alpha_{n,g} \gets \frac{\sum_{t \in T} p_{g,t}}{|T| \cdot P^{\max}_g}$ \\
        $k_{n+1,g} \gets k_{n,g} - m \cdot \alpha_{n,g}$
    }
}
\end{algorithm}

\section{Algorithm}\label{psudocode}
\Cref{algo:smr}. summarizes the formalism for rolling horizon short term dispatch of a nuclear reactor fleet with a reactivity tracking wrapper. Each scenario begins by initializing the fresh fuel multiplication factors, \(\mathrm{k_{\text{eff},BOL}}\), for individual reactors. The UC problem then determines the hourly generation of each reactor \( p_{g,t} \), using which the capacity factor \( \alpha_{0,g}\) for the initial period is computed. This feeds into the recursive relation in \Cref{eq:recursive_reln}, which updates the reactivity margin using \Cref{eq:margin} prior to the next dispatch window. The resulting value of \(\Delta \rho_{\text{marg},1} \) is used to adjust the minimum generation levels \( P_{g,1}^{\min} \) and downtime durations \( \text{MinDn}_{g,1} \) via the lookup tables. This iterative process is repeated for each period \( n \).

Once a reactor reaches refueling threshold (\( k_{\text{eff}} \simeq 1 \)), it undergoes an outage for a fixed duration \( \Delta t_{\text{refuel}} \). In this study, we assume \( \Delta t_{\text{refuel}} = 30 \) days or 1 month, consistent with the average refueling outage duration for large PWRs in the United States for 2024~\cite{noauthor_us_nodate}. During refueling, the reactor is offline and unavailable for generation. After the outage ends, the reactor restarts with a fresh fuel assembly, and the multiplication factor is reset to \( \mathrm{k_{\text{eff},BOL}} \). This sequence—dispatch, reactivity degradation, operational constraint update, and refueling—repeats across multiple fuel cycles, forming a closed loop feedback between economic dispatch decisions and nuclear physics constraints. The complete codebase, input datasets, and processed outputs for this study are available as open source on GitHub for community use \cite{shiny_shinychoudhuryphysics-informed-nuclear-reactor-unit-commitment-algorithm_2025}.

\section{Case Study}
\subsection{Basic Setup}
To demonstrate the proposed algorithm, we simulate the dispatch of a nuclear reactor fleet alongside a VRE-dominant mix that meets demand in the ERCOT South Central load zone~\cite{karg_bulnes_economic_2024}. Hourly demand data for the zone, as well as ERCOT-wide hourly wind and solar generation profiles for the years 2021–2023, were obtained from publicly available datasets~\cite{noauthor_grid_nodate}. Monthly installed capacities of wind and solar resources were used to compute hourly capacity factors for the VREs.

\begin{figure}
    \centering
    \includegraphics[scale=0.45]{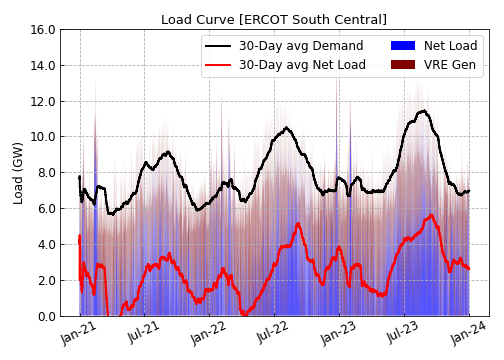}
    \caption{ERCOT South Central load and net-load statistics}
    \label{fig:ercot_south}
\end{figure}

A simplified capacity expansion model (CEM) was solved to determine the optimal installed capacities of wind, solar, and short-duration energy storage, assuming five AP1000 reactors as brownfield installations. The output capacities from the CEM serve as inputs for the modified UC model. The proposed framework is fully convex and readily extensible to include transmission constraints; however, in this study, network effects are omitted to preserve clarity of exposition. \Cref{tab:cem_output}. summarizes the optimal capacities obtained from the CEM and the cost assumptions used. The variable cost incorporates a heat rate of 
$10.44 \,\text{MMBtu/MWh}$, a fuel cost of \$0.50$\,$/MMBtu, and a variable operation and maintenance (O\&M) 
cost of \$4.55$\,$/MWh. Nuclear fleet cost parameters are based on the estimates provided by Stauff et al.~\cite{stauff_reactor_2021}. \Cref{fig:ercot_south}. presents the hourly demand, net load, and rolling average statistics to illustrate the temporal variability that the nuclear fleet must accommodate in the assumed load zone.

\begin{table}[ht!]
\centering
\caption{Input table for modified UC formalism}
\label{tab:cem_output}
\begin{tabular}{lccc}
\toprule
\textbf{Tech} & \textbf{Capacity} & \textbf{Variable} & \textbf{Start/Shut} \\
 & \textbf{(GW)} & \textbf{Cost (\$/MWh)} & \textbf{Cost (\$/MW)} \\
\midrule
Nuclear & 5.0 & 9.77 & 107.86 \\
Wind              & 8.0 & 0 & --- \\
Solar             & 12.0 & 0 & --- \\
Storage     & 6.0 (4-hr) & 0 & --- \\
\bottomrule
\end{tabular}
\end{table}

As mentioned earlier, nuclear reactors operate within a complex set of constraints that are exogenously set \cite{loflin_advanced_2014, stauff_reactor_2021}. And as a reactor progresses through its fuel cycle, additional constraints come into effect based on the reactivity margins. Consequently, the physics induced constraints are a function of the operational strategies and introduce a strong temporal coupling. In this study we endogenize the physics constraints and operate the reactors in different modes, as summarized in \Cref{tab:snr_description_table}. Mode-1 represents the traditional baseload operation of nuclear reactors, where the fleet can only startup or shutdown entire units to load follow. In contrast, Mode-2 and Mode-3 allow reactors to operate at reduced generation values, with hourly ramping levels constrained by predefined ramp up/down rates~\cite{stauff_reactor_2021}. These operational modes influence both the dispatchability of the fleet and the rate at which declining reactivity margin inflexibilities emerge, ultimately affecting the onset and frequency of refueling outages.

\begin{table}[htbp]
    \centering
    \caption{Modes of operation and parameter definition}
    \label{tab:snr_description_table}
    \begin{tabular}{lcc>{\raggedright\arraybackslash}p{2cm}}
        \toprule
         \textbf{Scenario} & \textbf{Min Gen}  & \textbf{Ramp Rate } & \textbf{Notes} \\
          \textbf{} & \textbf{(\%)}  & \textbf{(\%/hr)} & \textbf{} \\
        \midrule
         Mode-1  & 100\% & -- & Must run at full power or shutdown \\
         Mode-2  & 50\% & 25\% & Can ramp down to 50\% Min Gen \\
         Mode-3 & 20\%  & 25\% & Can ramp down to 20\% Min Gen \\
        \bottomrule
    \end{tabular}
\end{table}

\begin{figure*}
    \centering
    \begin{subfigure}{\linewidth}
        \centering
        \includegraphics[scale=0.43]{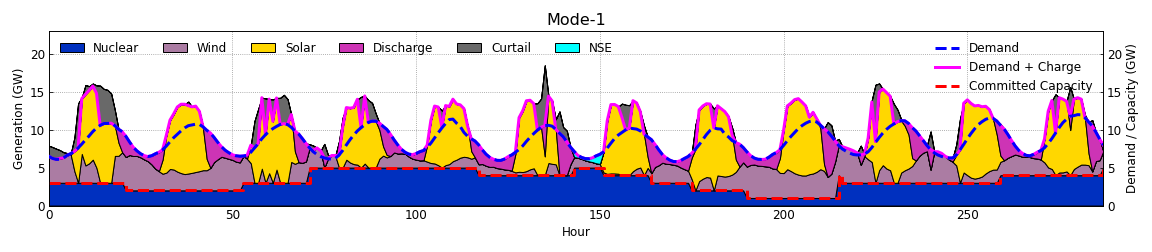}
        % \caption{GW\_MustRun}
    \end{subfigure}
    
    \begin{subfigure}{\linewidth}
        \centering
        \includegraphics[scale=0.43]{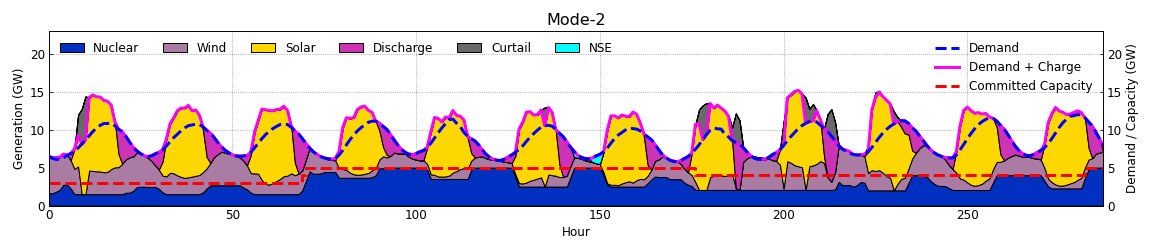}
        % \caption{GW\_PartLoad}
    \end{subfigure}
    
    \begin{subfigure}{\linewidth}
        \centering
        \includegraphics[scale=0.43]{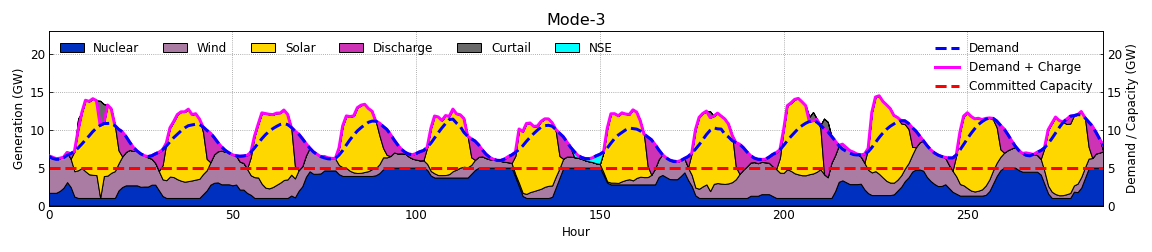}
        % \caption{GW\_PartLoad\_LowMinGen}
    \end{subfigure}

    \caption{12-day illustrative dispatch stackplot, demand, demand with charging and, committed capacity for different modes}
    \label{fig:three-dispatch}
\end{figure*}

The UC formalism is solved over a three-year dispatch horizon with rolling 3-day (72-hour) subproblems, and commitment states are transferred at subproblem boundaries. The choice of a 3-day window is made to ensure that the longest xenon induced downtime is contained within a single rolling horizon. The minimum stable time or \( \Delta t_{\text{stable}} \) is assumed to be 10 hours, which is equal to the base minimum downtime~\cite{jenkins_benefits_2018}. All simulations were performed on an Apple M1 MacBook with 64\,GB of RAM. 
The UC subproblem was formulated as a mixed-integer program and solved using the Gurobi solver with an academic license~\cite{gurobi}. An optimality gap tolerance of 0.1\% and a time limit of 1000 seconds were imposed per subproblem. To accelerate convergence, we enabled aggressive presolve and cut generation, employed the barrier
method for the root LP (with crossover disabled), and allowed parallel execution. The solver was configured to emphasize bound improvement (MIPFocus=1) with limited heuristic search (10\%), and detailed solver logs were retained for reproducibility.

\subsection{Results}
\begin{figure}
    \centering
    \includegraphics[scale=0.45]{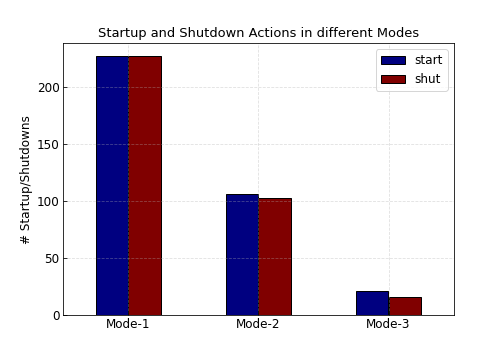}
    \caption{Mode dependent net startup/shutdown actions for year-one dispatch}
    \label{fig:start_shut}
\end{figure}

\Cref{tab:dispatchability_metrics}. summarizes key flexibility metrics including VRE curtailment, mean fleet capacity factor, production cost, and the time period for onset of refueling in different operational modes. Curtailment is lowest in Mode-3, where individual reactors can ramp down to minimal generation levels. This enhanced flexibility enables greater VRE integration. In contrast, Mode-1 exhibits the highest curtailment due to its restrictive operation: reactors may only startup or shutdown units entirely. This behavior is evident in the dispatch stackplot shown in \Cref{fig:three-dispatch}. Mode-1 incurs the highest production costs, driven by inflexible dispatch and the increased frequency of startup/shutdown actions. Without meaningful flexibility features, the fleet relies heavily on these costly actions. In Mode-3, however, the ability to sustain low generation levels allows reactor's to remain online, thereby lowering the need for expensive cycling and eventually reducing cost. The net startup/shutdown actions accumulated in each mode is shown in \Cref{fig:start_shut}. As evident, Mode-1 cycles its fleet excessively while, Mode-2 and -3 remain more conservative. Nuclear reactors are traditionally operated as baseload or in Mode-1. However, as indicated by the results, ramping abilities are crucial for greater VRE integration and economical operations.

\begin{table}[h]
    \centering
    \caption{Flexibility metrics for year-one of dispatch and onset of refueling period}
    \label{tab:dispatchability_metrics}
\begin{tabular}{ccccc>{\raggedright\arraybackslash}p{1.2cm}}
    \toprule
    \textbf{Scenario} & \textbf{VRE Curt} & \textbf{Fleet CF} & \textbf{Cost} & \textbf{Refuel } \\
    \textbf{} & \textbf{(\%)} & \textbf{(\%)} & \textbf{(mn\$)} & \textbf{ (year)} \\
    \midrule
    Mode-1 & 9.3\% & 56.82\% & 252 & $\sim$ 2.55   \\
    Mode-2 & 8.1\% & 55.03\% & 216 & $\sim$ 2.65 \\
    Mode-3 & 6.4\% & 53.19\% & 188 & $\sim$ 2.8 \\
    \bottomrule
\end{tabular}
\end{table}

\begin{figure*}
    \centering
    \begin{subfigure}{\linewidth}
        \centering
        \includegraphics[scale=0.43]{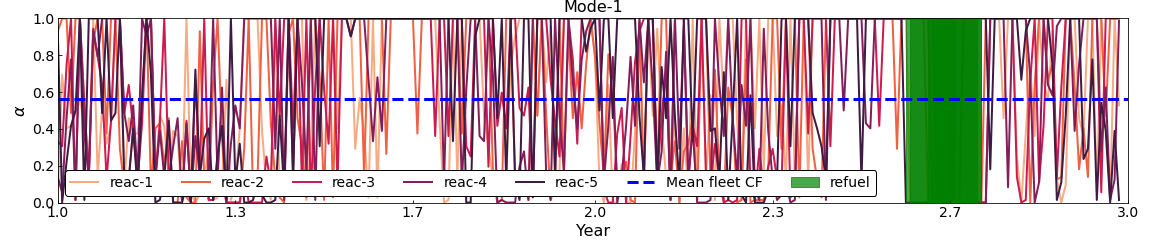}
        % \caption{GW\_MustRun}
    \end{subfigure}
    
    \begin{subfigure}{\linewidth}
        \centering
        \includegraphics[scale=0.43]{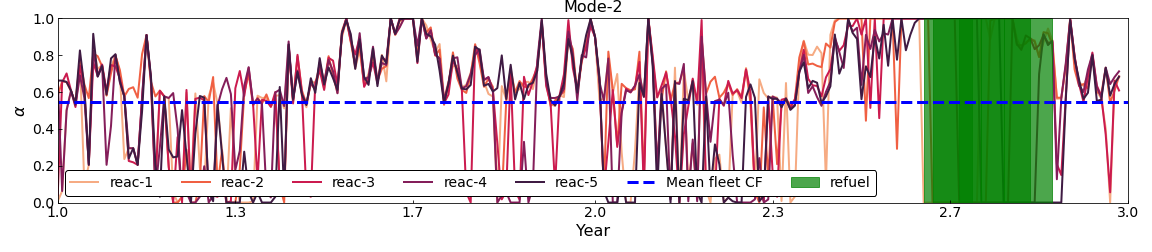}
        % \caption{GW\_PartLoad}
    \end{subfigure}
    
    \begin{subfigure}{\linewidth}
        \centering
        \includegraphics[scale=0.43]{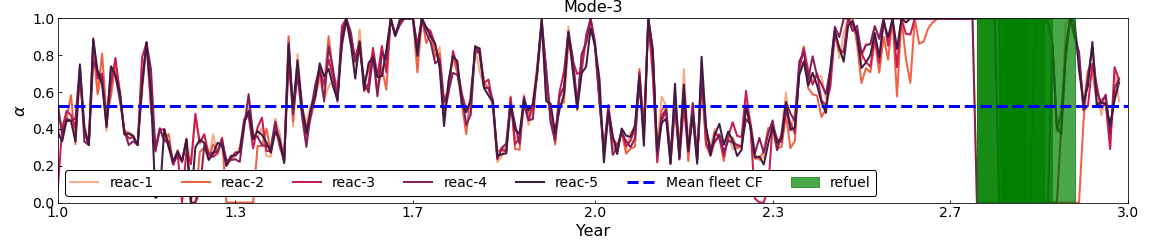}
        % \caption{GW\_PartLoad\_LowMinGen}
    \end{subfigure}

    \caption{$\alpha$ evolution of nuclear fleet in different operational modes year-two onwards}
    \label{fig:keff_evolution}
\end{figure*}

\Cref{fig:keff_evolution}. shows the capacity factor ($\alpha$) evolution for individual reactors across the dispatch period. The plot highlights key differences in how fleets achieve flexibility under different modes. In Mode-1, the fleet cycles individual reactors based on their commitment statuses. Reactor downtimes are long and costly but unavoidable in this mode, since startup and shutdown are the only available flexibility levers. During periods of high net-load variability, some reactors operate at low capacity factors while others remain online for the entire horizon to maximize dispatchability. In Modes-2 and -3, fewer startup/shutdown actions combined with access to lower generation levels allow the fleet to remain online albeit at lower capacity factors. But as fleets approach critical reactivity margins, downtimes elongate futher and minimum generation levels rise. This drives reactors to remain online and generating at full power despite faster fuel burnup and higher VRE curtailemnt. 

\Cref{tab:dispatchability_metrics}. also reports the average period before onset of refueling in different modes. Mode-1 reaches refueling the fastest due to its higher average fleet capacity factor. In contrast, operating reactors flexibly allows fleets to achieve longer refueling cycles due to lower utilization. Interestingly, in all three modes we see significant elongation of average refueling periods due stratergic cycling between -low and -high reactivity margin reactors. These observations are consistent with findings by Alhadhrami et al., who reported that flexible and staggered operation extends refueling intervals \cite{alhadhrami_dispatch_2023}. Our formalism captures this behavior by adapting refueling dynamics to the operational mode. 

Across the different modes, reactors behave identically nearing refueling, as all flexibility levers are declining. Individual reactors lose flexibility as end-of-cycle conditions arise reducing operational maneuverability. Within the fleet, reactors can still cycle between more flexible and less flexible units, delaying individual refueling events and maximizing fleet-level flexibility. However, once a reactor reaches $P_{\min} = 100\%$, it offers no flexibility. Even shutdown actions are not optimal due to long downtimes. At this stage decisions are a techno-economic trade-off, where higher burnup may be preferred if it minimizes overall production costs. Altogether, these results demonstrate that while operational modes govern short-term dispatch, they also have long-term system level implications through reactivity margins and the irreducible physics of nuclear reactors.

\section{Conclusion}
Most energy system models represent nuclear reactors with fixed, inflexible operational constraints, thereby underestimating their potential flexibility. In reality, a reactor’s constraints are tightly coupled to its fuel cycle and associated reactivity margins. For much of the cycle, reactors can respond flexibly to grid signals; however, towards the end, core physics imposes additional binding constraints. These stem from the buildup of fission products, gradual degradation of control rod worth, and xenon poisoning—particularly following a power ramp down, which suppresses core reactivity and induces power stagnation. To capture this, we develop lookup tables that encode maneuverability limits by increasing minimum generation levels and enforcing extended downtimes as a function of the reactor’s core state. These tables are used to augment the canonical unit commitment constraints, allowing for a more accurate representation of nuclear reactor operation. In summary, this study introduces a novel formulation that integrates reactivity degradation, core state dependent operational constraints alongside refueling and economic dispatch within a unit commitment framework. Using this framework, we dispatch a fleet of large nuclear reactors in three distinct operational modes within a VRE-heavy energy mix. Mode-1 reflects traditional baseload operation, where the fleet can only follow load by starting up or shutting down entire units. In contrast, Mode-2 and Mode-3 enable part load operation, allowing reactors to reduce power adhering to predefined ramp rates. We observe that, flexible modes reduce both VRE curtailment and production costs. While Mode-1 results in the highest production cost due to accumulation of cycling costs due to significant startup/shutdown actions used by the fleet. Moreover, we find that flexible operation, as in Mode-2 and Mode-3, slows reactivity degradation and delays both xenon-induced inflexibilities and the onset of refueling outages. With more flexible reactor operation, refueling outages are expected to become longer and more distributed over time. However, since reactor scheduling is primarily driven by economic factors, commitment and dispatch decisions will ultimately depend on the market and energy mix. In particular, the renewed demand growth driven by AI and data center loads may incentivize more flexible nuclear operation—especially when capital costs are absorbed by data center operators deploying nuclear alongside VRE. In such settings, nuclear flexibility is likely to be actively utilized to enhance system reliability and support extended refueling cycles. This work contributes a generalized nuclear reactor representation that maximizes fleet-level flexibility in energy system models, while offering a physics-informed, computationally tractable formalism that embeds both regulatory constraints and core physics into operational scheduling.

\bibliographystyle{IEEEtran}
\bibliography{zotero}

\begin{thebibliography}{10}
\providecommand{\url}[1]{#1}
\csname url@samestyle\endcsname
\providecommand{\newblock}{\relax}
\providecommand{\bibinfo}[2]{#2}
\providecommand{\BIBentrySTDinterwordspacing}{\spaceskip=0pt\relax}
\providecommand{\BIBentryALTinterwordstretchfactor}{4}
\providecommand{\BIBentryALTinterwordspacing}{\spaceskip=\fontdimen2\font plus
\BIBentryALTinterwordstretchfactor\fontdimen3\font minus \fontdimen4\font\relax}
\providecommand{\BIBforeignlanguage}[2]{{%
\expandafter\ifx\csname l@#1\endcsname\relax
\typeout{** WARNING: IEEEtran.bst: No hyphenation pattern has been}%
\typeout{** loaded for the language `#1'. Using the pattern for}%
\typeout{** the default language instead.}%
\else
\language=\csname l@#1\endcsname
\fi
#2}}
\providecommand{\BIBdecl}{\relax}
\BIBdecl

\bibitem{haratyk_early_2017}
\BIBentryALTinterwordspacing
G.~Haratyk, ``\BIBforeignlanguage{en}{Early nuclear retirements in deregulated {U}.{S}. markets: {Causes}, implications and policy options},'' \emph{\BIBforeignlanguage{en}{Energy Policy}}, vol. 110, pp. 150--166, Nov. 2017. [Online]. Available: \url{https://linkinghub.elsevier.com/retrieve/pii/S0301421517305177}
\BIBentrySTDinterwordspacing

\bibitem{impram_challenges_2020}
\BIBentryALTinterwordspacing
S.~Impram, S.~Varbak~Nese, and B.~Oral, ``Challenges of renewable energy penetration on power system flexibility: {A} survey,'' \emph{Energy Strategy Reviews}, vol.~31, p. 100539, Sep. 2020. [Online]. Available: \url{https://www.sciencedirect.com/science/article/pii/S2211467X20300924}
\BIBentrySTDinterwordspacing

\bibitem{schill_start-up_2017}
\BIBentryALTinterwordspacing
W.-P. Schill, M.~Pahle, and C.~Gambardella, ``\BIBforeignlanguage{en}{Start-up costs of thermal power plants in markets with increasing shares of variable renewable generation},'' \emph{\BIBforeignlanguage{en}{Nature Energy}}, vol.~2, no.~6, pp. 1--6, Apr. 2017, number: 6 Publisher: Nature Publishing Group. [Online]. Available: \url{https://www.nature.com/articles/nenergy201750}
\BIBentrySTDinterwordspacing

\bibitem{gonzalez-salazar_review_2018}
\BIBentryALTinterwordspacing
M.~A. Gonzalez-Salazar, T.~Kirsten, and L.~Prchlik, ``Review of the operational flexibility and emissions of gas- and coal-fired power plants in a future with growing renewables,'' \emph{Renewable and Sustainable Energy Reviews}, vol.~82, pp. 1497--1513, Feb. 2018. [Online]. Available: \url{https://www.sciencedirect.com/science/article/pii/S1364032117309206}
\BIBentrySTDinterwordspacing

\bibitem{shaker_impacts_2016}
\BIBentryALTinterwordspacing
H.~Shaker, H.~Zareipour, and D.~Wood, ``Impacts of large-scale wind and solar power integration on {California}'s net electrical load,'' \emph{Renewable and Sustainable Energy Reviews}, vol.~58, pp. 761--774, May 2016. [Online]. Available: \url{https://www.sciencedirect.com/science/article/pii/S1364032115016706}
\BIBentrySTDinterwordspacing

\bibitem{lund_review_2015}
\BIBentryALTinterwordspacing
P.~D. Lund, J.~Lindgren, J.~Mikkola, and J.~Salpakari, ``Review of energy system flexibility measures to enable high levels of variable renewable electricity,'' \emph{Renewable and Sustainable Energy Reviews}, vol.~45, pp. 785--807, May 2015. [Online]. Available: \url{https://www.sciencedirect.com/science/article/pii/S1364032115000672}
\BIBentrySTDinterwordspacing

\bibitem{jenkins_benefits_2018}
\BIBentryALTinterwordspacing
J.~Jenkins, Z.~Zhou, R.~Ponciroli, R.~Vilim, F.~Ganda, F.~de~Sisternes, and A.~Botterud, ``\BIBforeignlanguage{en}{The benefits of nuclear flexibility in power system operations with renewable energy},'' \emph{\BIBforeignlanguage{en}{Applied Energy}}, vol. 222, pp. 872--884, Jul. 2018. [Online]. Available: \url{https://linkinghub.elsevier.com/retrieve/pii/S0306261918303180}
\BIBentrySTDinterwordspacing

\bibitem{stauff_reactor_2021}
\BIBentryALTinterwordspacing
N.~Stauff, W.~N. Mann, K.~Biegel, T.~Levin, J.~D. Rader, A.~Cuadra, and S.~H. Kim, ``\BIBforeignlanguage{English}{Reactor {Power} {Size} {Impacts} on {Nuclear} {Competitiveness} in a {Carbon}-{Constrained} {Future}},'' Argonne National Lab. (ANL), Argonne, IL (United States), Tech. Rep. ANL/NSE-21/55, Sep. 2021. [Online]. Available: \url{https://www.osti.gov/biblio/1865640}
\BIBentrySTDinterwordspacing

\bibitem{franceschini_advanced_2008}
\BIBentryALTinterwordspacing
F.~Franceschini and B.~Petrovic, ``\BIBforeignlanguage{en}{Advanced operational strategy for the {IRIS} reactor: {Load} follow through mechanical shim ({MSHIM})},'' \emph{\BIBforeignlanguage{en}{Nuclear Engineering and Design}}, vol. 238, no.~12, pp. 3240--3252, Dec. 2008. [Online]. Available: \url{https://www.sciencedirect.com/science/article/pii/S002954930800294X}
\BIBentrySTDinterwordspacing

\bibitem{lamarsh_introduction_2001}
J.~R. Lamarsh and A.~J. Baratta, \emph{\BIBforeignlanguage{en}{Introduction to nuclear engineering}}, 3rd~ed., ser. Addison-{Wesley} series in nuclear science and engineering.\hskip 1em plus 0.5em minus 0.4em\relax Upper Saddle River, N.J: Prentice Hall, 2001.

\bibitem{ponciroli_profitability_2017}
\BIBentryALTinterwordspacing
R.~Ponciroli, Y.~Wang, Z.~Zhou, A.~Botterud, J.~Jenkins, R.~B. Vilim, and F.~Ganda, ``\BIBforeignlanguage{en}{Profitability {Evaluation} of {Load}-{Following} {Nuclear} {Units} with {Physics}-{Induced} {Operational} {Constraints}},'' \emph{\BIBforeignlanguage{en}{Nuclear Technology}}, vol. 200, no.~3, pp. 189--207, Dec. 2017. [Online]. Available: \url{https://www.tandfonline.com/doi/full/10.1080/00295450.2017.1388668}
\BIBentrySTDinterwordspacing

\bibitem{rahman_steady-state_2025}
\BIBentryALTinterwordspacing
J.~Rahman and J.~Zhang, ``\BIBforeignlanguage{en}{Steady-{State} {Modeling} of {Small} {Modular} {Reactors} for {Multi}-{Timescale} {Power} {System} {Operations} {With} {Temporally} {Coupled} {Sub}-{Models}},'' \emph{\BIBforeignlanguage{en}{IEEE Transactions on Power Systems}}, vol.~40, no.~1, pp. 793--805, Jan. 2025. [Online]. Available: \url{https://ieeexplore.ieee.org/document/10522977/}
\BIBentrySTDinterwordspacing

\bibitem{lynch_nuclear_2022}
\BIBentryALTinterwordspacing
A.~Lynch, Y.~Perez, S.~Gabriel, and G.~Mathonniere, ``Nuclear fleet flexibility: {Modeling} and impacts on power systems with renewable energy,'' \emph{Applied Energy}, vol. 314, p. 118903, May 2022. [Online]. Available: \url{https://www.sciencedirect.com/science/article/pii/S0306261922003282}
\BIBentrySTDinterwordspacing

\bibitem{alhadhrami_dispatch_2023}
\BIBentryALTinterwordspacing
S.~Alhadhrami, G.~J. Soto, and B.~Lindley, ``Dispatch analysis of flexible power operation with multi-unit small modular reactors,'' \emph{Energy}, vol. 280, p. 128107, Oct. 2023. [Online]. Available: \url{https://www.sciencedirect.com/science/article/pii/S0360544223015013}
\BIBentrySTDinterwordspacing

\bibitem{loflin_advanced_2014}
\BIBentryALTinterwordspacing
L.~Loflin and B.~McRimmon, ``\BIBforeignlanguage{English}{Advanced {Nuclear} {Technology}: {Advanced} {Light} {Water} {Reactors} {Utility} {Requirements} {Document} {Small} {Modular} {Reactors} {Inclusion} {Summary}},'' Electric Power Research Institute, Inc., Knoxville, TN (United States), Tech. Rep. 3002003130, Dec. 2014. [Online]. Available: \url{https://www.osti.gov/biblio/1165570}
\BIBentrySTDinterwordspacing

\bibitem{noauthor_ap1000_nodate}
\BIBentryALTinterwordspacing
``{AP1000} {ARIS} {Specifications}.'' [Online]. Available: \url{https://aris.iaea.org/PDF/AP1000.pdf}
\BIBentrySTDinterwordspacing

\bibitem{ingersoll_can_2015}
D.~Ingersoll, C.~Colbert, Z.~Houghton, R.~Snuggerud, J.~Gaston, and M.~Empey, ``Can {Nuclear} {Power} and {Renewables} be {Friends}?'' in \emph{Proceedings of ICAPP}, vol.~9, 2015, p.~19.

\bibitem{knueven_mixed-integer_2020}
\BIBentryALTinterwordspacing
B.~Knueven, J.~Ostrowski, and J.-P. Watson, ``\BIBforeignlanguage{en}{On {Mixed}-{Integer} {Programming} {Formulations} for the {Unit} {Commitment} {Problem}},'' \emph{\BIBforeignlanguage{en}{INFORMS Journal on Computing}}, p. ijoc.2019.0944, Jun. 2020. [Online]. Available: \url{http://pubsonline.informs.org/doi/10.1287/ijoc.2019.0944}
\BIBentrySTDinterwordspacing

\bibitem{carrion_computationally_2006}
\BIBentryALTinterwordspacing
M.~Carrion and J.~Arroyo, ``A computationally efficient mixed-integer linear formulation for the thermal unit commitment problem,'' \emph{IEEE Transactions on Power Systems}, vol.~21, no.~3, pp. 1371--1378, Aug. 2006. [Online]. Available: \url{https://ieeexplore.ieee.org/document/1664974}
\BIBentrySTDinterwordspacing

\bibitem{noauthor_us_nodate}
\BIBentryALTinterwordspacing
``U.{S}. summer nuclear outages declined in 2024, returning to 2022 levels - {U}.{S}. {Energy} {Information} {Administration} ({EIA}).'' [Online]. Available: \url{https://www.eia.gov/todayinenergy/detail.php?id=63624}
\BIBentrySTDinterwordspacing

\bibitem{shiny_shinychoudhuryphysics-informed-nuclear-reactor-unit-commitment-algorithm_2025}
\BIBentryALTinterwordspacing
S.~Choudhury, ``shinychoudhury/physics-informed-nuclear-reactor-unit-commitment-algorithm: v1,'' Aug. 2025. [Online]. Available: \url{https://zenodo.org/records/16971060}
\BIBentrySTDinterwordspacing

\bibitem{karg_bulnes_economic_2024}
\BIBentryALTinterwordspacing
F.~Karg~Bulnes, D.~Hofer, N.~R. Smith, J.~Schmitt, O.~Pryor, G.~Khawly, and A.~McClung, ``\BIBforeignlanguage{en}{Economic {Modeling} of {Renewable} {Integration} and the {Application} of {Energy} {Storage} to the {ERCOT} {Energy} {Grid}},'' in \emph{\BIBforeignlanguage{en}{Volume 6: {Education}; {Electric} {Power}; {Energy} {Storage}; {Fans} and {Blowers}}}.\hskip 1em plus 0.5em minus 0.4em\relax London, United Kingdom: American Society of Mechanical Engineers, Jun. 2024, p. V006T08A003. [Online]. Available: \url{https://asmedigitalcollection.asme.org/GT/proceedings/GT2024/87981/V006T08A003/1204098}
\BIBentrySTDinterwordspacing

\bibitem{noauthor_grid_nodate}
\BIBentryALTinterwordspacing
``Grid {Information}.'' [Online]. Available: \url{https://www.ercot.com/gridinfo}
\BIBentrySTDinterwordspacing

\bibitem{gurobi}
\BIBentryALTinterwordspacing
{Gurobi Optimization, LLC}, ``{Gurobi Optimizer Reference Manual},'' 2024. [Online]. Available: \url{https://www.gurobi.com}
\BIBentrySTDinterwordspacing

\end{thebibliography}

\appendix
% \section*{Appendix}

\begin{table}[!thb]
\centering
\caption{Key Parameters for Westinghouse AP1000}
\label{tab:parameter}
\begin{adjustbox}{width=\columnwidth}
\begin{tabular}{|l|c|}
\hline
\textbf{Parameter} & \textbf{Value} \\
\hline
% Nuclear unit electrical output, $P_0$ & 1000 MW$_{\text{electric}}$ \\
% Operational range, $(P_{\min}, P_{\max})$ & (15\% ; 100\%) \\
% Max. power rate of change, $(dP/dt)_{\max}$ & $\pm$5\%/min \\
Average neutron flux, $\phi_0$ & $1.8 \times 10^{17}$ n/cm$^2$·hr \\
Macroscopic fission cross-section, $\bar{\Sigma}_f$ & 0.39497 cm$^{-1}$ \\
Iodine decay constant, $\lambda_I$ & 0.01033 hr$^{-1}$ \\
Xenon decay constant, $\lambda_{Xe}$ & 0.0753 hr$^{-1}$ \\
Iodine effective yield, $\gamma_I$ & 0.0639 \\
Xenon effective yield, $\gamma_{Xe}$ & 0.00237 \\
Xenon microscopic absorption cross-section, $\sigma_{abs}^{Xe}$ & $2.65 \times 10^{-18}$ cm$^2$ \\
Neutrons per fission, $\nu$ & 2.42 \\
\hline
\end{tabular}
\end{adjustbox}
\end{table}

\end{document}